\documentclass[aps,prl,reprint,showpacs]{revtex4-1}
\usepackage{hyperref}
\usepackage{amsmath}
\usepackage{braket}
\usepackage{natbib}
\usepackage{graphicx}

\newcommand{\eqaref}[1]{Eq.\eqref{#1}}

\begin{document}
\title{Quantised orbital angular momentum transfer and magnetic dichroism in the interaction of electron vortices with matter}
\author{Sophia Lloyd}
\author{Mohamed Babiker}
\author{Jun Yuan}
\affiliation{Deaprtment of Physics, University of York, Heslington, York, YO10 5DD, UK}
\pacs{41.85-p, 42.50Tx}

\begin{abstract}
Following the very recent experimental realisation of electron vortices, we consider their interaction with matter, in particular the transfer of orbital angular momentum in the context of electron energy loss spectroscopy, and the recently observed dichroism in thin film magnetised iron samples.  We show here that orbital angular momentum exchange does indeed occur between electron vortices and the internal electronic-type motion, as well as center of mass motion of atoms in the electric dipole approximation.  This contrasts with the case of optical vortices where such transfer only occurs in transitions involving multipoles higher than the dipole. The physical basis of the observed dichroism is explained. 
\end{abstract}

\maketitle

The suggestion for the existence of electron vortices (EVs) was first put forward by Bliokh \textit{et al}.~\cite{Bliokh2007}. In analogy with optical vortices (OVs) \cite{Allen1999, Allen2003, Andrews2008, Grier2003}, EVs are also endowed with the property of quantized orbital angular momentum (OAM) of $l\hbar$ per electron, where $l$ is the winding number, but differ from OVs in that EVs have electric charge, mass, and spin of $\frac{1}{2}$. Bliokh \textit{et al}.'s highly significant prediction was followed by experimental work which succeeded in the generation of EVs,  beginning with the first experiments by Uchida and Tonomura \cite{Uchida2010}, who used a stepped spiral phase plate, followed by the work of Verbeeck \textit{et al}.~\cite{Verbeeck2010}, who used a binary holographic grating with a Y-like point defect. More recently, using this holographic method,  McMorran \textit{et al}.~have shown that it is possible to generate beams with winding numbers as high as $l=100$ \cite{McMorran2011}. It is now clear that EVs are easily generated in conventional electron microscopes with the help of holographic reproduction techniques.  The potential applications are both diverse and wide-ranging and it is now known that EVs are set to revolutionize electron microscopy and spectroscopy, promising markedly improved sub-nanometre resolution in the imaging of materials, including biological specimens of materials with low absorption contrast \cite{Jesacher2005}.   

As shown theoretically by Berry \cite{Berry1998a} and demonstrated amply in the context of OVs \cite{Allen1999, Allen2003, Andrews2008, Grier2003}, it is clearly reasonable to state that the orbital angular momentum is a well defined quantised property of EVs and hence should be exchanged when an electron vortex beam interacts with an atom, a molecule, or a solid; for instance, in electron energy loss spectroscopy  (EELS) using electron vortex beams. Furthermore, one expects OAM to be exchanged not just in electric dipole interactions, but also in electric quadrupole and higher electric multipole interactions involving the bound state internal `electronic-type' motion, as well as the gross `center of mass-type' motion.  However, in an electric dipole transition the exchange of OAM between OVs and matter has been shown \cite{Babiker2002} to affect only the center of mass-type motion, so that to this leading multipolar order no OAM transfer occurs from OVs  to the internal `electronic' motion. This prediction has been confirmed experimentally \cite{Araoka2005, Loffler2011}.

This article reports the results of investigations seeking to explore the exchange of OAM between EVs and matter and to highlight the respects in which EVs differ from OVs in the processes of exchange of OAM in this context.  We also seek to explain the results of recent experimental studies by Verbeeck \textit{et al}., in which the observation of dichroism is reported  in electron energy loss spectroscopy of thin film magnetized iron samples using EVs  \cite{Verbeeck2010}.  

For simplicity we concentrate on the most basic model for matter, displaying both internal `electronic-type' bound states relative to the center of mass, as well as gross motion of the center of mass.  The model in question is a two-particle hydrogenic atom interacting with the electron vortex beam.  The EV beam is also taken in the simplest form, namely a Bessel beam, which is one of the possible solutions of the paraxial Schr\"{o}dinger equation, having well defined orbital angular momentum about the beam axis.  The interaction between the beam electron and the atom is taken as the leading Coulomb interaction.  For the total Hamiltonian, we write
\begin{equation}
\hat{H}= \hat{H}_{0}^{e}+\hat{H}_{0}^{N}+\hat{H}_{0}^{V}+\hat{H}_{\text{int}},
\end{equation}
where $\hat{H}_{0}^{e}$ is the Hamiltonian operator for the unperturbed internal electron motion relative to the center of mass;  its eigenstates are written as $\ket{\psi_{e}}\equiv\ket{n;\ell;m}$; the well known hydrogenic states such that
\begin{equation}
\hat{H}_{0}^{e}\ket{n;\ell;m}=E_{n}\ket{n;\ell;m},
\end{equation}
where the symbols have their usual meanings. The second term $\hat{H}_{0}^{N}$ is the Hamiltonian of the unperturbed 
center of mass motion, with eigenstates written $\ket{\psi_{N}}\equiv\ket{\mathbf{K}; L}$   describing both translational and rotational motions of the center of mass;
\begin{equation}
{\hat H}_{0}^{N}\ket{\mathbf{K}; L}=(E_{K}+E_{R})\ket{\mathbf{K}; L},
\end{equation}
where $L$ stands for angular momentum of the center of mass about the beam axis; $E_K$ and $E_R$ are the energies associated with the translational and rotational motion respectively.  The eigenfunctions of the center of mass are products of plane waves of wavevector $\mathbf{K}$ - proportional to $\exp{(i\mathbf{K\cdot R})}$ where $\mathbf{R}$ is the center of mass position variable - and rotational states proportional to $\exp{(iL\Phi_{R})}$ where $\Phi_{R}$ is the azimuthal angle of the center of mass in cylindrical coordinates relative to the center of the electron vortex  beam. For the unperturbed EV, the ${\hat H}_{0}^{V}$ has eigenfunctions written as  $\ket{\psi_{B}}\equiv \ket{k_z;k_{\rho};l}$ such that
\begin{equation}
{\hat H}_{0}^{V}\ket{k_z;k_{\rho};l}=(E_{k}+E_{\rho})\ket{k_z;k_{\rho};l},
\end{equation}
where the total energy consists of both translational axial motion of axial wavevector $k_{z}$ and rotational motion about the beam axis.  The explicit eigenstates are given in cylindrical coordinates in the form
\begin{equation}
\ket{\psi_{B}(\rho,\Phi_r,z)}=\ket{k_z;k_{\rho};l}=A_{l}J_{l}(k_{\rho}\rho)e^{il\Phi_{r}}e^{ik_{z}z},
\label{vortex}
\end{equation}
where $A_{l}$ is a normalisation factor and $k_{\rho}$ is the radial component of the wavevector, such that total wavevector $k^2=k_z^2+k_{\rho}^2$.   Finally $\hat{H}_{\text{int}}$ is the interaction Hamiltonian which, to leading order, is given by the Coulomb interaction energy
\begin{equation}
\hat{H}_{\text{int}}=e_{0}^{2}\left(\frac{1}{|{\mathbf{r}-\mathbf{q}_{e}}|}-\frac{1}{|{\mathbf{r}-\mathbf{R}}|}\right),
\label{coulomb}
\end{equation}
where $e_{0}^2=e^{2}/(4\pi\epsilon_{0})$ and $\mathbf{q}_{e}$ defines the coordinates of the atomic electron relative to the center of the beam. In writing \eqaref{coulomb}, we have made the usual simplification that the nucleus is much more massive than the electron so that the center of mass is essentially located at the nuclear coordinate, $\mathbf{R}$.  Note that, in the above, the azimuthal angles denoted by the capital $\Phi$ are measured with respect to the cylindrical coordinate frame centered about the electron beam.  Azimuthal angles denoted by $\phi$ are measured with respect to the spherical coordinate frame positioned at the nuclear center of mass.
\begin{figure}%
\includegraphics[width=\columnwidth]{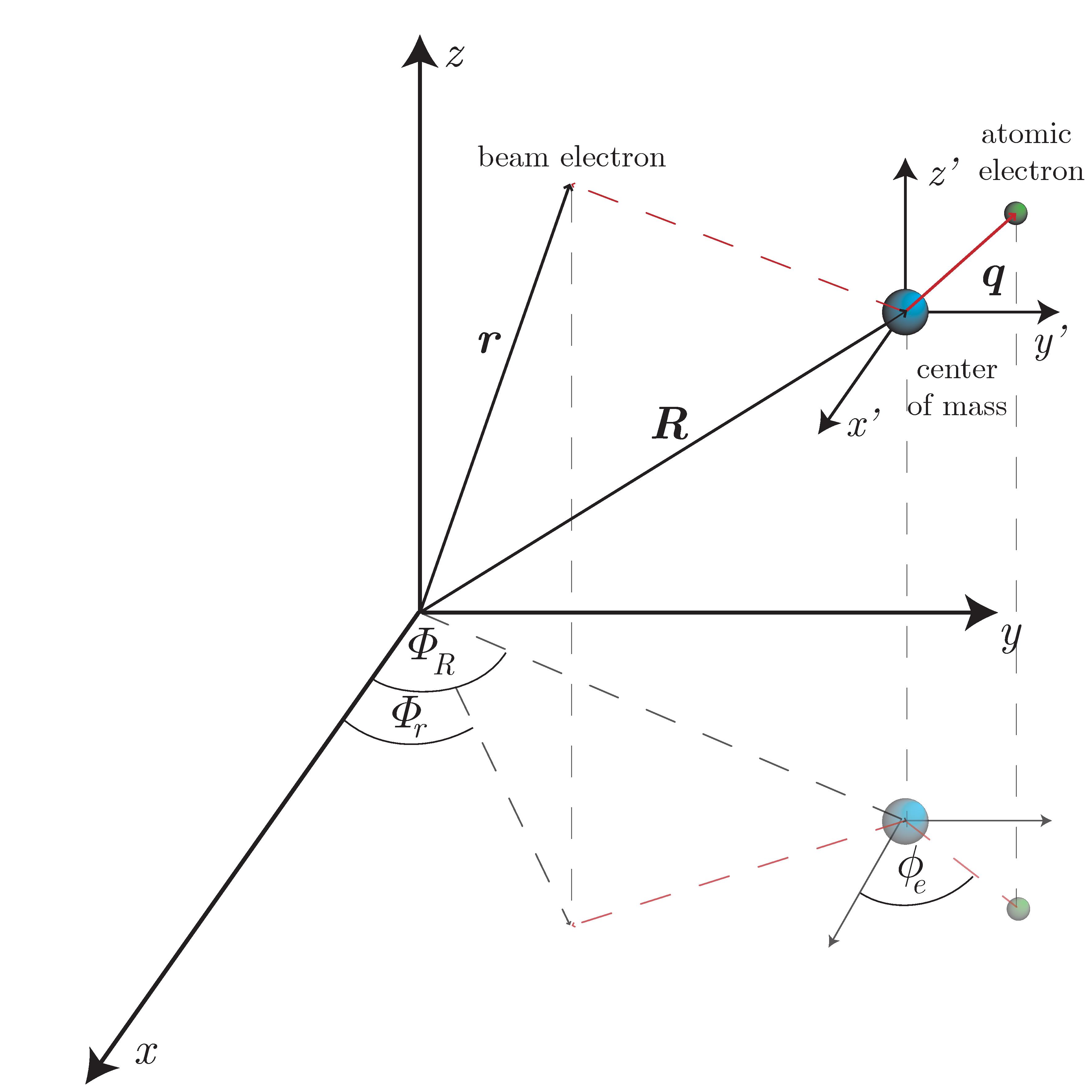}%
\caption{The relevant coordinate frames in the description of the interaction between a Bessel electron vortex and a two-particle neutral system.  Position within the beam, $\mathbf{r}$, is described in cylindrical coordinates about the frame $xyz$; the center of mass of the atom is located at $\mathbf{R}$, and the atomic electron is descibed by $\mathbf{q}$ in spherical coordinates relative to the center of mass.  Projection in the $xy$ plane indicates azimuthal angles $\Phi_r$, $\Phi_R$, and $\phi_e$ of the beam, atomic center of mass and atomic electron respectively.}%
\label{Coordinate system}%
\end{figure}

The evaluation of effects arising from the interaction between the EV state (henceforth to be referred to as the beam electron) with the two-particle atom requires consideration of the transition matrix element $\mathcal{M}_{if}$ where $\ket{i}\equiv \ket{\psi_{e};\psi_{N};\psi_{B}}$ and $\ket{f}\equiv \ket{\psi_{e}';\psi_{N}';\psi_{B}'}$. We have
\begin{equation}
\mathcal{M}_{if}=\braket{\psi_{e}';\psi_{N}';\psi_{B}'|{\hat H}_{int}|\psi_{e};\psi_{N};\psi_{B}}.
\label{matrixelement}
\end{equation} 
An important technical problem we have to consider here is that the atomic electron wavefunctions are normally given in spherical polar coordinates relative to the atomic center of mass $\mathbf{R}$, whilst in \eqaref{coulomb} both the center of mass and the atomic electron are expressed relative to the center of the EV beam, using cylindrical polar coordinates.  We are interested in evaluating the transition matrix element in such a manner that information can be gained as to whether, due to the interaction, a transfer of orbital angular momentum has occurred between the three subsystems; namely the EV beam, the atomic center of mass and the atomic electron. Since the transitions between internal atomic states are classified as electric dipole-active, electric quadrupole-active, etc., we seek to determine the transfer selection rules in the different multipolar approximations. We wish to find out whether an experiment can be performed to detect OAM exchange between the electron vortex beam and the internal dynamics of an atomic or molecular system through changes involving electric dipole transitions.  It turns out, and this constitutes one of the main findings of this Letter, that in contrast to the case of optical vortex interaction with such systems, as discussed in \cite{Babiker2002}, the transfer of OAM from electron vortices is possible in the electric dipole approximation.  We shall therefore restrict ourselves to this approximation in the interaction Hamiltonian, \eqaref{coulomb}. We have
\begin{equation}
\hat{H}_{\text{int}}\approx e_{0}^{2}\frac{\mathbf{q}\cdot (\mathbf{r}-\mathbf{R})}{|\mathbf{r}-\mathbf{R}|^{3}},
\label{dipolepot}
\end{equation}
where $\mathbf{q}=(\mathbf{q}_{e}-\mathbf{R})$ is the internal electron coordinate relative to the center of mass, such that $e\mathbf{q}$ is the electric dipole moment.  

To proceed we compute explicitly the scalar product in \eqaref{dipolepot}.  With reference to Fig.\ref{Coordinate system}, and after simplification, the interaction Hamiltonian can  be written as
\begin{equation}
\hat{H}_{int}\approx e_{0}^{2}\frac{q\chi_{r}\cos(\phi_{e}-\Phi_{r})-q\chi_{R}\cos(\phi_{e}-\Phi_{R})+q\eta}{\left[\mathcal{F}-\mathcal{G}\cos(\Phi_{r}-\Phi_{R})\right]^{\frac{3}{2}}},
\label{hint2}
\end{equation} 
where $\chi_{r}$, $\chi_{R}$, $\eta$, $\mathcal{F}$, and $\mathcal{G}$ are functions independent of the electron-center of mass coordinate $q$.  In the next step towards evaluating the matrix element \eqaref{matrixelement}, we evaluate the integral over azimuthal angles.  We write
\begin{multline}
\mathcal{M}_{\text{azimuth}} \propto \int e^{i(l-l')\Phi_{r}}e^{i(L-L')\Phi_{R}}\\
\times e^{i(m-m')\phi_{e}}\hat{H}_{int}d\Phi_{r}d\Phi_{R}d\phi_{e},
\end{multline}
where $m$ and $m'$ are magnetic quantum numbers associated with the transition between atomic states $|n;\ell;m>$ and $|n';\ell';m'>$. In evaluating this integral we make the substitution $y=\Phi_{r}-\Phi_{R}$, in order to isolate the integral over $\Phi_r$.  We find that this integral can be expressed in terms of generic integrals of the form
\begin{equation}
Y_{\alpha}=\int_{0}^{2\pi}\frac{e^{i(l-l'+\alpha)y}}{[\mathcal{F}-\mathcal{G}\cos{(y)}]^{3/2}}dy,
\end{equation} 
where $\alpha$ is an integer. The final result emerging from the angular integral can be written as 
\begin{multline}
\mathcal{M}_{\text{azimuth}} \propto \mathcal{Q} \ \delta_{[(l+L),(l'+L'+1)]}\delta_{m,m'-1}\\
{}+\mathcal{S} \ \delta_{[(l+L),(l'+L'-1)]}\delta_{m,m'+1}\\
{}+\mathcal{U} \ \delta_{[(l+L),(l'+L')]}\delta_{m,m'},
\label{mefinal}
\end{multline}
where $\mathcal{Q}$, $\mathcal{S}$ and $\mathcal{U}$ are complex functions of coordinates and are all first order in the internal atomic electron coordinate $\mathbf{q}$, consistent with electric dipole transitions. In particular, it turns out that 
\begin{equation}
\mathcal{Q}^{*}=\mathcal{S}.
\end{equation}
The interpretation of the result in \eqaref{mefinal} is as follows.  The first term indicates a dipole active transition in which the EV beam or the atomic center of mass loses one unit of orbital angular momentum due to a upwards transition in which the atom gains one unit of orbital angular momentum.  The second term indicates that the one unit transfer occurs from the atomic electron to either the EV beam or the center of mass.  The final term shows that transitions can occur in which transfer of orbital angular momentum occurs only between the vortex and the center of mass rotation, and the internal electronic motion is not involved.  In situations where the atomic center of mass is rigidly fixed, as in a solid, so that $L=L'$,  the exchange of orbital angular momentum would occur between the EV and the atomic electron.  Thus we have shown that the electric dipole transition can mediate the transfer of a single unit of orbital angular momentum between the vortex and the internal motion of the atomic electron.  Similarly we can show that higher multipole transitions lead to exchange of two or more units of orbital angular momentum between the vortex beam and the atomic system, but this will not be further discussed here.  

The difference in dipole selection rules between OVs and EVs - that absorption is forbidden for an OV and allowed by an EV - is interesting, as it suggests that an OAM electron beam could be more useful in the context of spectroscopy, as the dipole transition is often the dominant process in most physical systems.  For example, it means that EV beams could be used to detect circular dichroic activity in biological molecules such as proteins, in order to gain information about their secondary structures \cite{Greenfield1969, Chen1972}.  If radiation damage effects can be mitigated, the sensitivity and spatial resolution will be high.

The selection rules obtained above form the basis for the explanation of  the origin of the dichroic signal observed in the L$_{2}$ and L$_{3}$ magnetized iron thin film  by Verbeeck \textit{et al}.~\cite{Verbeeck2010}.  As only the $m$-selection rule is of importance in this single particle transition matrix element, the same selection rules will emerge if we replace the single particle wavefunction of the hydrogenic atom in the above example by the many-particle wavefunctions of a transition metal atom, i.e. replacing $\ket{i}=\ket{n;l;m}$ by $\ket{2p^6 3d^n;j;m_j}$, and $\ket{f}=\ket{n';l';m'}$ by $\ket{2p^5 3d^{n+1};j';m_j'}$, where $j$ and $m_j$ are the orbital angular momentum and magnetic quantum numbers characterizing the many-particle configuration consisting of the participating $2p$ core electrons and $3d$ valence electrons in the transition metal atom \cite{Thole1992}.

Using the matrix elements defined above we first consider the azimuthal angular dependence of the transition rate $\Gamma_{l}$ for transition induced by an EV having $l=\pm1$, using Fermi's golden rule,
\begin{equation}
\Gamma_{l}\propto|\mathcal{M}_{\text{azimuth}}|^2\rho_{f},
\label{Fermi}
\end{equation}
where $\rho_{f}$ is the density of final states.  Once again only the angular factor is needed for our purpose, as we are interested in the effect of the beam's orbital angular momentum on the transition.  The relevant transitions are the $m'= m\pm 1$ transitions, and in order for these transitions to occur, the one unit of angular momentum of $l=\pm1$ must be gained or lost by the electron vortex.  We therefore seek to compare the transition rates for  $l$ and $-l$.  Consider first the dipole transition $m'= m+1$, $l=+1$.  We find from \eqaref{mefinal}
\begin{equation}
\mathcal{M}^{l=+1}_{\text{azimuth}} \propto \mathcal{Q},
\end{equation}
while for $m'= m-1$, $l=-1$ we have from 
\begin{equation}
\mathcal{M}^{l=-1}_{\text{azimuth}} \propto -\mathcal{S}=-\mathcal{Q}^{*},
\label{negativel}
\end{equation}
where the same proportionality factor applies in the two cases.  This proportionality factor includes the radial integrals, involving  Bessel functions arising from the initial and final beam states, and since $J_{-1}(x)=-J_{+1}(x)$  a factor of $-1$ arises in \eqaref{negativel}.  It thus follows, since the transition rate depends on the modulus square of the matrix element which is the same for $l=\pm 1$, and provided that the densities of final magnetic states remain the same, i.e.~$\tilde{\rho}_{m+1}=\tilde{\rho}_{m-1}$, we will have
\begin{equation}
\Gamma_{l}=\Gamma_{-l}.
\end{equation}
This means there is no difference in the strength between the two transitions involving different helicities $l=\pm 1$.  There is therefore no dichroism for non-magnetic transitions.  

The clear dichroism observed in the experiment by Verbeeck \textit{et al}.~\cite{Verbeeck2010} can thus be explained as arising due to the difference in the density of states $\tilde{\rho}_f$, which will differ for each $m'$.  The fact that it is a dipole mediated transition means that the dichroic signal can be measured using small angle scattering, where such dipole interactions dominate.  This has clear advantages compared to the electron energy-loss magnetic chiral dichroism (EMCD) technique devised by Schattschneider \textit{et al}.~\cite{Schattschneider2006, Schattschneider2008} in which large scattering angles are required.

In conclusion, we have shown by direct analysis that it is possible to transfer OAM between an electron vortex beam and the internal electron states of an atom in the dipole transition and we have checked by direct analysis that OAM transfer occurs for quadupole transitions and in principle in the case of all higher multipoles.  This is in direct contrast to the case of optical OAM transfer in the interaction with similar systems.  It has been demonstrated both theoretically and experimentally that optical vortices are not specific in their interaction with chiral matter. Here we have shown that although orbital angular momentum transfer can occur between electron vortices and matter in electric dipole transitions for a given topological charge $|l|$, there is no intrinsic difference in absorption for the two opposite helicities $\pm l$.  We have concluded that the OAM dichroic electron energy-loss spectroscopy of the type performed by Verbeeck \textit{et al}.~\cite{Verbeeck2010} shows a dichroism due to the magnetic nature of the material, in which magnetic sublevels would be unequally populated.

The authors are grateful to J.~Verbeeck for useful discussion.

\bibliography{References}{}
\end{document}